\documentclass[aps,prl,showpacs,twocolumn,superscriptaddress,floatfix]{revtex4-1}

\usepackage{graphicx}
\usepackage{color}
\usepackage[colorlinks,bookmarks=false,citecolor=blue,linkcolor=red,urlcolor=blue]{hyperref}
\usepackage{bm}
\usepackage{amsmath,amssymb}

\newcommand{\la}{\langle}
\newcommand{\ra}{\rangle}
\newcommand{\bS}{\mathbf{S}}
\newcommand{\dg}{\dagger}

\begin{document}

\title{Carrier driven coupling in ferromagnetic oxide heterostructures}

\author{Ching-Hao Chang}
\affiliation{Leibniz-Institute for Solid State and Materials Research, IFW-Dresden, D-01171 Dresden, Germany }
\author{Sujit Das}
\affiliation{Institute for Physics, MLU Halle-Wittenberg, 06099 Halle, Germany}
\affiliation{Leibniz-Institute for Solid State and Materials Research, IFW-Dresden, D-01171 Dresden, Germany }
\author{Sanjeev Kumar}
\affiliation{Indian Institute of Science Education and Research Mohali, Sector 81, S.A.S. Nagar, Manauli PO 140306, India }
\author{R. Ganesh}
\affiliation{Leibniz-Institute for Solid State and Materials Research, IFW-Dresden, D-01171 Dresden, Germany }
\affiliation{The Institute of Mathematical Sciences, C I T Campus, Chennai 600 113, India}
\email{ganesh@imsc.res.in}

\date{\today}

\begin{abstract}
Transition metal oxides are well known for their complex magnetic and electrical properties. 
When brought together in heterostructure geometries, they show particular promise for spintronics and colossal magnetoresistance applications.  
In this letter, we propose a new mechanism for the coupling between layers of itinerant ferromagnetic materials in heterostructures. 
The coupling is mediated by charge carriers that strive to maximally delocalize through the heterostructure to gain kinetic energy. In doing so, they force a ferromagnetic or antiferromagnetic coupling between the constituent layers. To illustrate this, we focus on heterostructures composed of SrRuO$_3$ and La$_{1-x}$A$_{x}$MnO$_3$ (A=Ca/Sr).
Our mechanism is consistent with antiferromagnetic alignment that is known to occur in multilayers of SrRuO$_3$-La$_{1-x}$A$_{x}$MnO$_3$.
To support our assertion, we present a minimal Kondo-lattice model which reproduces the known magnetization properties of such multilayers.
In addition, we discuss a quantum well model for heterostructures and argue that the spin-dependent density of states determines the nature of the coupling. 
As a smoking gun signature, we propose that bilayers with the same constituents will oscillate between ferromagnetic and antiferromagnetic coupling upon tuning the relative thicknesses of the layers. 
\end{abstract}


\maketitle

\paragraph{Introduction:} Magnetic heterostructures have been intensively studied driven by the technological promise of spintronics\cite{Bibes2007} and colossal magnetoresistance\cite{Ramirez1997}. Their interesting properties derive from the new physics that emerges when two different magnetic materials come into contact at an interface\cite{Hwang2012}. It is generally assumed that it is the interface that decides the coupling between layers and the overall properties of the heterostructure. For instance, ferromagnetic/antiferromagnetic alignments are usually attributed to superexchange interactions across the interface.
In this letter, we demonstrate a new coupling mechanism driven by conduction electrons that diffuse deep into each constituent layer. Macroscopic behaviour is determined not by the interface, but by the nature of charge carriers. Our study paves the way for tailoring spintronics devices with strongly coupled magnetic and transport properties.

\begin{figure}
\includegraphics[width=3in]{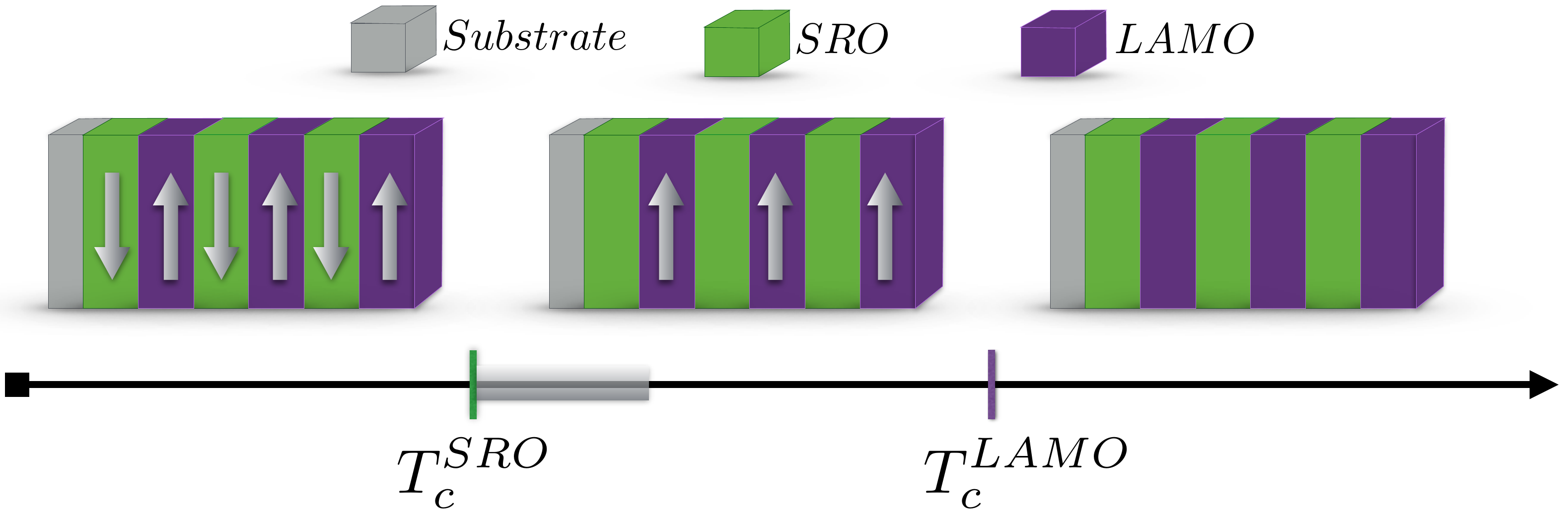}
\caption{Phase diagram of an SRO-LAMO multilayer heterostructure. The shaded regime above $T_c^{SRO}$ has incipient ferromagnetic correlations in the SRO layers.}
\label{fig.Tdiagram}
\end{figure}

To illustrate this new coupling, we place our discussion within the context of heterostructures with alternating layers of SrRuO$_3$ (SRO) and La$_{1-x}$A$_{x}$MnO$_3$ (LAMO) with $x\sim0.3$ and A=Ca/Sr. 
Both LAMO and SRO are layered metallic ferromagnets. While LAMO is a conventional double exchange system, SRO is a bad metal in which conduction is dominated by minority-spin electrons\cite{Koster2012,Worledge2000}. 
Multilayer SRO/LAMO heterostructures have been extensively studied with a view to understand the interlayer coupling.
The typical $M$-$T$ phase diagram has two sharp features\cite{Ziese2010}. Upon approaching from high temperatures, the LAMO layers first develop ferromagnetic order at $T_c^{LAMO}\sim 300K$. Upon further cooling, the SRO layers order in the opposite direction  at $T_c^{SRO}\sim 150 K$ and the net moment starts decreasing. The $M$-$h$ curve shows three hysteresis 
loops\cite{Ziese2010,Solignac2012}, with the SRO layer flipping its ordered moment at some intermediate field. Taken together, these effects demonstrate the antiferromagnetic coupling between the layers with a phase diagram as shown in Fig.~\ref{fig.Tdiagram}. Density functional calculations performed at zero temperature also suggest antiferromagnetic coupling, attributing it to the role of interfacial Oxygen atoms\cite{Ziese2010}.
Bilayers of SRO and LAMO have also been studied, with several studies finding antiferromagnetic coupling\cite{Solignac2012,SolignacIEEE2012,Borisevich2012}. 

We propose that the inter-layer coupling in these systems is driven by charge carriers and not by interfacial Mn-O-Ru bonds. The key ingredient in our proposal is the minority carrier nature of SRO: 
its conduction electrons are predominantly polarized \textit{opposite} to the direction of its ferromagnetic moment. On the other hand, LAMO, being a double exchange ferromagnet, has carriers which are polarized \textit{parallel} to the ferromagnetic moment. 
An antiferromagnetic alignment of the ordered moments allows the conduction electrons to polarize the same way in both materials. Electrons can then maximally delocalize over the heterostructure and gain kinetic energy.  
This mechanism is clearly consistent with the antiferromagnetic ordering seen in multilayer heterostructures. In contrast, bilayers present a more nuanced situation depending on the density of states. We show that the carrier-driven mechanism predicts that the coupling will oscillate with the relative widths of layers. 

\textit{Minimal model for multilayer geometries}: 
As a minimal model for multilayer heterostructures, we present
a Kondo-lattice model of local moments coupled to itinerant electrons. 
Our objective is to show that the itinerancy of conduction electrons suffices to explain the nature of interlayer coupling, with no need for interfacial defects or interfacial superexchange couplings. 

\begin{figure}
\includegraphics[width=3in]{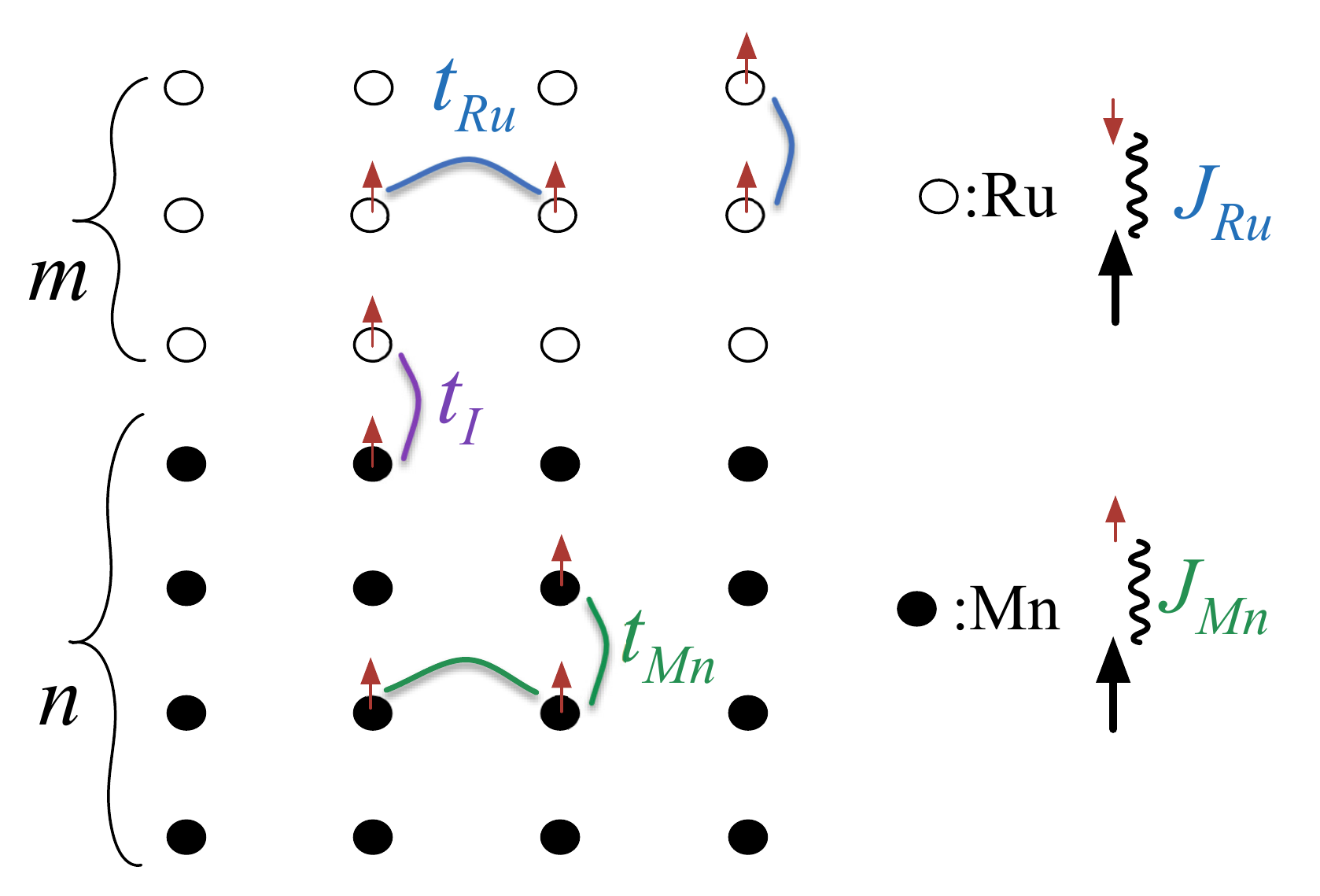}
\caption{A schematic view of the lattice model for $n$-LAMO/$m$-SRO heterostructures. There are three different hopping parameters, $t_{Mn}$, $t_{Ru}$ and $t_{I}$, and two different Kondo couplings,
$J_{Mn}$ and $J_{Ru}$.}
\label{fig.model_lattitce}
\end{figure}

Our model is composed of square lattice divided into two regions, one with Mn and one with Ru ions, as shown in Fig.~\ref{fig.model_lattitce}. 
Note that Mn and Ru are the magnetically active ions in LAMO and SRO respectively. 
At each site, we have a local moment $\bS_i$ as well as an orbital that may be occupied by a conduction electron. 
The spin of the conduction electron at a given site is coupled to that of the corresponding local moment. The resulting Hamiltonian is given by
\begin{eqnarray}
H &=& \sum_{i,\sigma,\sigma'} J_{i} \bS_i \cdot \{ \frac{1}{2}c_{i,\sigma}^\dg \vec{\tau}_{\sigma,\sigma'} c_{i,\sigma'}  \} - \sum_{\la ij \ra,\sigma} \{ t_{ij} c_{i,\sigma}^\dg c_{j,\sigma}  + h.c.\} \nonumber \\
&+ & \!\!\sum_{i,\sigma} (\epsilon_i - \mu) c_{i,\sigma}^\dg c_{i,\sigma} \!-\! h\sum_{i} \{ S_i^z \!+\! \frac{1}{2}\sum_{\sigma,\sigma'}c_{i,\sigma}^\dagger \tau_{\sigma,\sigma'}^z c_{i,\sigma'}\!\}.
\label{eq.Hamiltonian}
\end{eqnarray}
 The operator $c_{i,\sigma}^\dg$ ($c_{i,\sigma}$) creates (annihilates) an itinerant electron at site $i$ with spin $\sigma$. The components of $\vec{\tau}$ are Pauli matrices representing spin operators. We have included an external magnetic field $h$. 
The parameters $J_i$ and $t_{ij}$ take different values within each material. The materials are coupled by hopping across the interface, with strength $t_{I}$. The local potential $\epsilon_i$ takes the value $\Delta$($0$) for Mn(Ru) sites. The relative on-site potential  $\Delta$ is adjusted to achieve desired average electronic density on SRO and LAMO sides. 

\paragraph{Parameters appropriate for LAMO-SRO multilayers:}
We model the ($n$-LAMO/$m$-SRO) multilayer as a cluster with $m$ atomic layers of Ru and $n$ atomic layers of Mn with overall periodic boundary conditions. Taking $t_{Mn} = 1$, we choose $t_{Ru}=0.5$ to capture the relatively low value of $T_c$ for SRO compared to LAMO. Along the interface, we fix the hopping $t_{I}$ to be $0.5$. 
The chemical potential $\mu$, and the relative on-site potential $\Delta$ are fixed so as to give an average electronic filling of $0.7$ ($0.5$) for LAMO (SRO). 
We find the same qualitative results irrespective of the specific values of the parameters. 

In SRO, the magnetic moments originate from Ru$^{4+}$ (4d$^4$) ions residing in an octahedral crystal field.
\textit{Ab initio} calculations show that the valence electrons occupy t$_{2g}$ orbitals, while the e$_g$ orbitals remain empty\cite{Koster2012, Mahadevan2009,Jeng2006}.
In the atomic limit, the four electrons in 
the t$_{2g}$ levels form a high-spin state with a total moment of $2\mu_B$/Ru atom. If two such atoms are brought together, the minority spins can hop between atoms whereas the majority 
spin carriers will be localized due to Pauli blocking. As several atoms are brought together to form the material, itinerancy effects reduce the total moment on each Ru site to $\sim 1.5 \mu_B$/Ru atom\cite{Koster2012}. 
We model the Ru site as a $2$ $\mu_B$ local moment coupled to itinerant electrons, with an average filling of 0.5 per site. The Kondo coupling is taken to be positive ($J_{Ru} > 0$) to account for the minority carrier nature.
DFT+DMFT studies also point to a dual nature of magnetism in SRO consisting of a Stoner behaviour as well as 
local magnetic moments above $T_{c}$ \cite{Jakobi2011,Kim2015,Dang2015}.
 We note that a realistic description of SRO requires a multi-orbital model with easy axis anisotropy, spin-orbit coupling, etc. However, our simple one-orbital model captures the essential aspect required for the carrier-driven mechanism, which is the minority carrier character.

In LAMO, on the other hand, the Mn ions are nominally in the $+3$ ($3d^{4}$) state.  Three electrons with parallel spins occupy the t$_{2g}$ levels, while the fourth electron occupies an e$_g$ orbital. 
In undoped LaMnO$_3$, superexchange mechanism drives the net S=2 spins to order antiferromagnetically (the e$_g$ orbital degree of freedom also orders). Replacing a fraction of the La atoms by Sr or Ca, 
removes electrons from the e$_g$ levels and gives rise to the classic double exchange scenario. The filled t$_{2g}$ electrons form a S=3/2 local moment. The e$_g$ electrons can maximally delocalize when 
the local moments order ferromagnetically. With this picture in mind, we assume a local moment of $3$ $\mu_B$/Mn site. 
The Kondo coupling is negative ($J_{Mn} < 0$) reflecting the standard double exchange scenario. This scenario is well established in the context of bulk manganites \cite{Dagotto1998a,Hotta2003}. In particular, in the optimally doped regime, $x \sim 0.3$, a single-band ferromagnetic Kondo lattice model can describe the magnetic and transport properties very well \cite{Millis1995,Roder1996,Kumar2006a}. We fix $J_{Mn}$ to be $-20$ and present results for different values of $J_{Ru}$.

\begin{figure}
\includegraphics[width=3.25in]{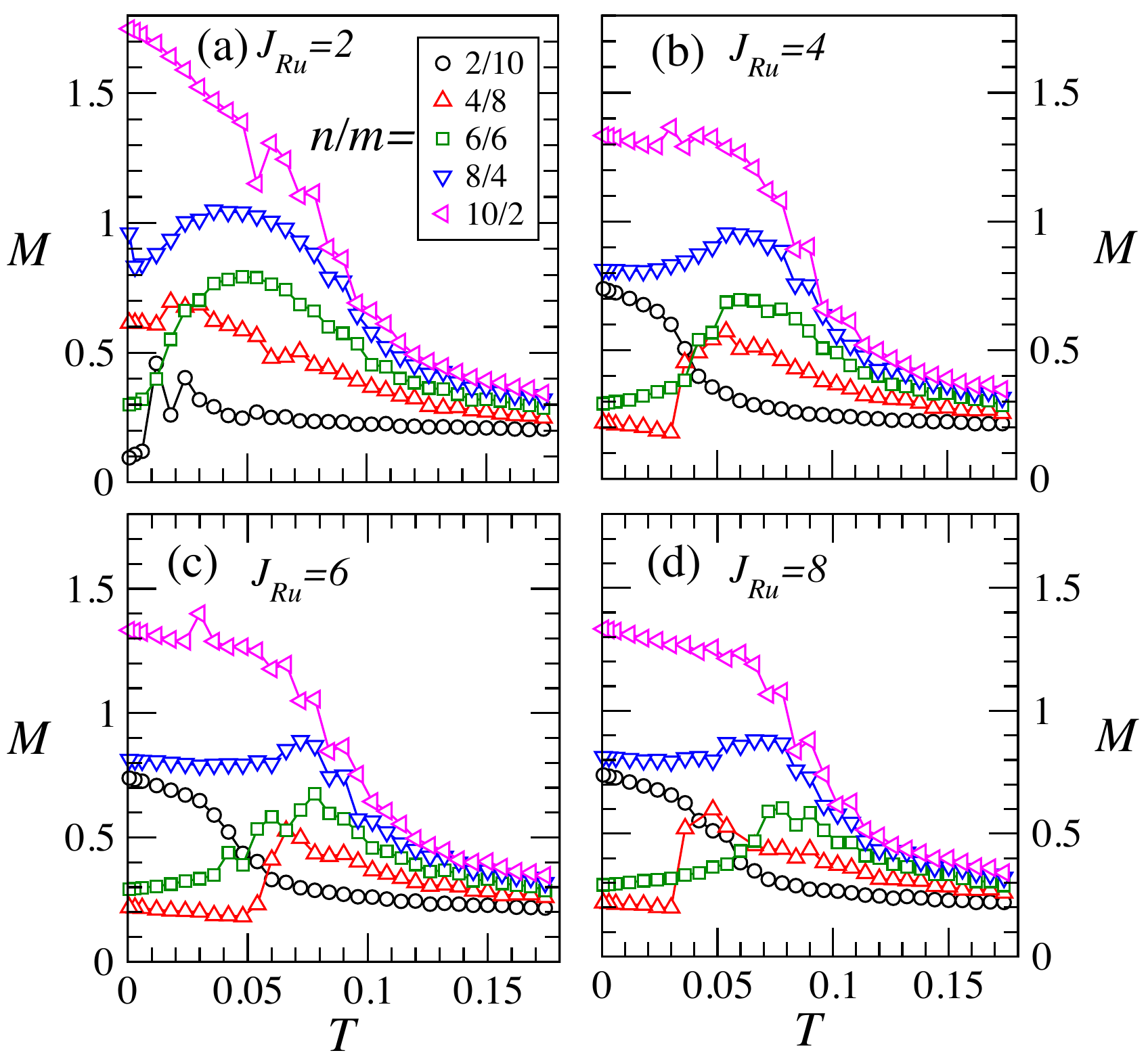}
\caption{ (a)-(d) Magnetization as a function of temperature for different values of $J_{Ru}$. Each panel shows results for different relative thicknesses of LAMO-SRO layers as indicated by
$n/m$. $n$ and $m$ are the number of LAMO and SRO rows in the unit cell respectively, with $(n+m)=12$ fixed. The results are obtained on a 4$\times$12 lattice with periodic boundary conditions. The value of $J_{Mn}$ is fixed to be $-20$.
} 
\label{fig.M-T_simulations}
\end{figure}

\paragraph{Semi-classical Monte Carlo simulations:}
To solve for the properties of the model, we use a protocol involving exact diagonalization of the conduction electrons combined with classical Monte Carlo sampling of the local moment configurations (see Supplementary Materials for details)\cite{Dagotto2001a}. 
Unlike ab initio calculations, this method is ideally suited to explore finite temperature properties. These computations are CPU intensive and therefore restricted to small clusters;
however, they allow for an unbiased exploration of magnetic configuration space. We present results for the magnetization, defined as $\frac{1}{N_{av}}\sum_{\alpha} \vert M_{\alpha} \vert$, 
where $N_{av}$ is the number of Monte Carlo configurations used for averaging. $M_{\alpha}$ is the magnetization for each configuration, defined as,
\begin{equation}
M_{\alpha} = \frac{1}{N} \sum_{i} ({\bS}_i + \la \vec\tau_i \ra ),
\end{equation}
where, $N$ is the total number of sites and angular brackets denote the quantum expectation values. Our results for magnetization vs. temperature are plotted in Fig.~\ref{fig.M-T_simulations}, for different values of $J_{Ru}$ and different relative widths of SRO and LAMO layers.
As long as $J_{Ru} \geq 4$, we find the same qualitative behaviour. For example, for a given choice of layer widths, the $T \rightarrow 0$ values of total magnetization are identical for $J_{Ru} = 4, 6, 8$.

\begin{figure}
\includegraphics[width=3.25in]{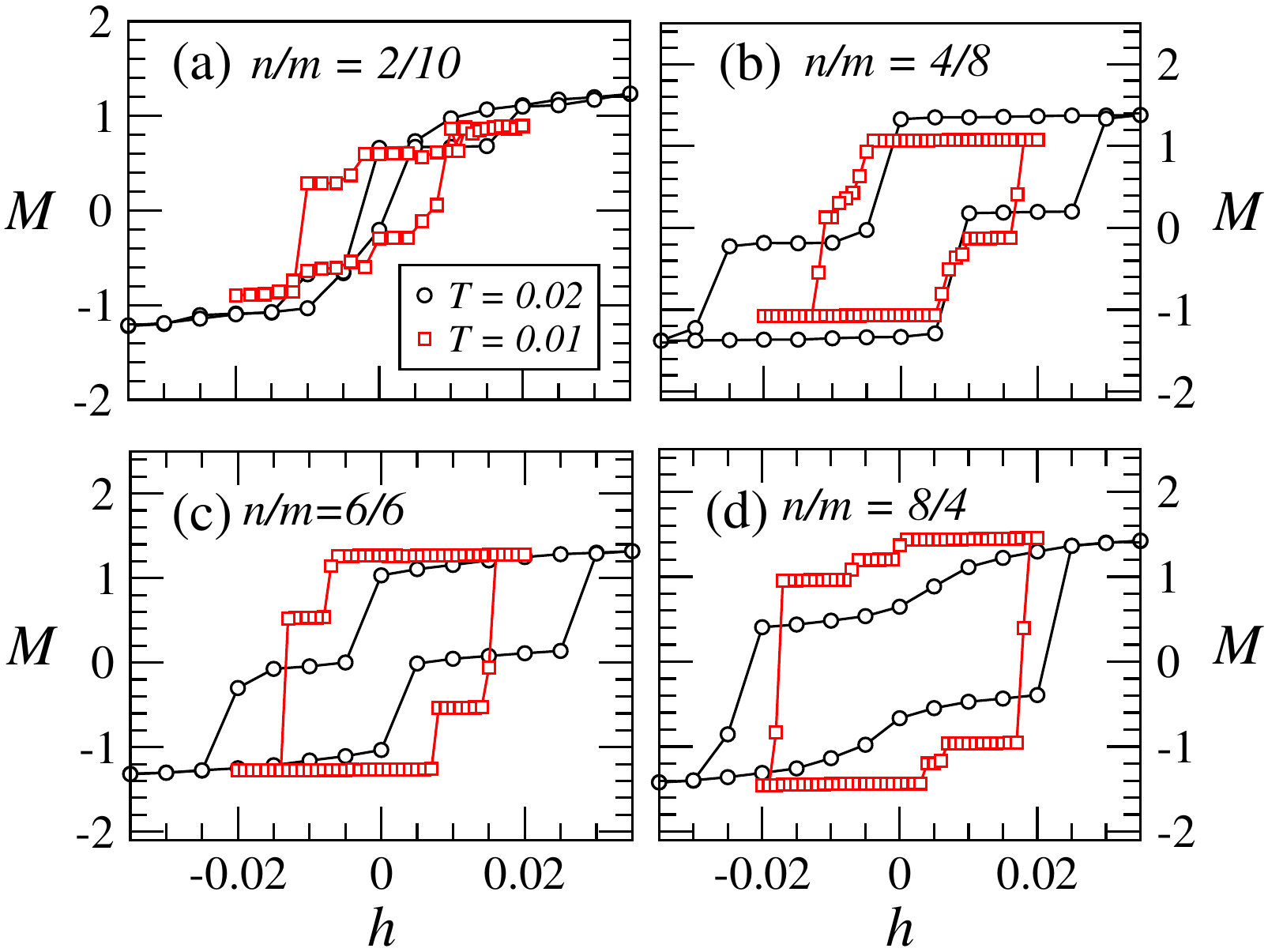}
\caption{ Magnetization as a function of applied magnetic field with $J_{Mn}=-20$, $J_{Ru} = 4$, $t_I = 0.5$.
Results are obtained on 4$\times 12$ lattices, with $n$ rows of LAMO rows and $m$ rows of SRO, with $(n+m)=12$. } 
\label{fig.M-H_simulations}
\end{figure}

While our simulations on small clusters cannot capture thermodynamic phase transitions, they can show features that are indicative of phase transitions. The inflection points in the magnetization curve provide a first estimate of $T_c$. This can be taken as the temperature below which the correlation length exceeds the cluster size. For LAMO, we find $T_c \sim 0.1 t_{Mn}$. SRO shows paramagnetic behaviour until $T \lesssim 0.04 t_{Mn}$, below which the SRO cluster also orders ferromagnetically.
The reduction of total magnetization indicates that the SRO ferromagnetic moment is oriented opposite to that of LAMO. This behaviour is in good agreement with experimental
results\cite{Ziese2010}. We further support this picture by computing $M$-$h$ hysteresis curves at low temperatures as shown in Fig.~\ref{fig.M-H_simulations}. The double-loop feature in the $M$-$h$ curves also agrees well with experiments\cite{Ziese2010,Solignac2012}. It reflects a two-step switching process as a function of $h$, i.e., LAMO layers align along the field first, while SRO layers align at a stronger field. 

\paragraph{Quantum well model:}
We rationalize our results for the inter-layer coupling in multilayers within a quantum well picture; we will later extend these arguments to make predictions for bilayers. 
To understand the inter-layer coupling, we focus on temperatures immediately above  $T_c^{SRO}\sim 150 K$. In this regime, the LAMO layers have already developed ferromagnetic order; we assume that their magnetic moment has formed along the `up' direction. The conduction electrons inside the LAMO layers are spin polarized with spins pointing up, as LAMO is a conventional `majority-carrier' ferromagnet. Within our minimal quantum well picture, we assume maximum spin polarization -- i.e., down-spin electrons are not present inside the LAMO layers.

\begin{figure}
\includegraphics[width=3.5in]{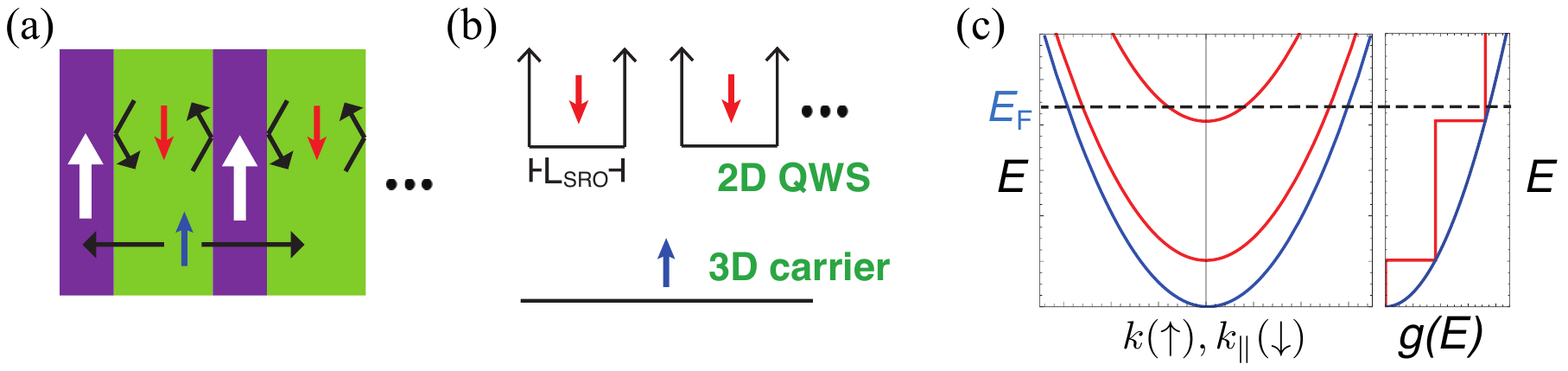}
\caption{Quantum well description of SRO-LAMO multilayers: (a) Multilayer geometry, showing spin polarization just above $T_c^{SRO}$, assuming that the LAMO layers have ordered along the `up' direction. (b) Potentials seen by the two spin species; the up electrons can diffuse throughout while the down electrons are confined to the SRO layers. (c) The resulting band energies and DOS for the two spin species.  }
\label{fig.QW_multilayer}
\end{figure}

\begin{figure}
\includegraphics[width=3.in]{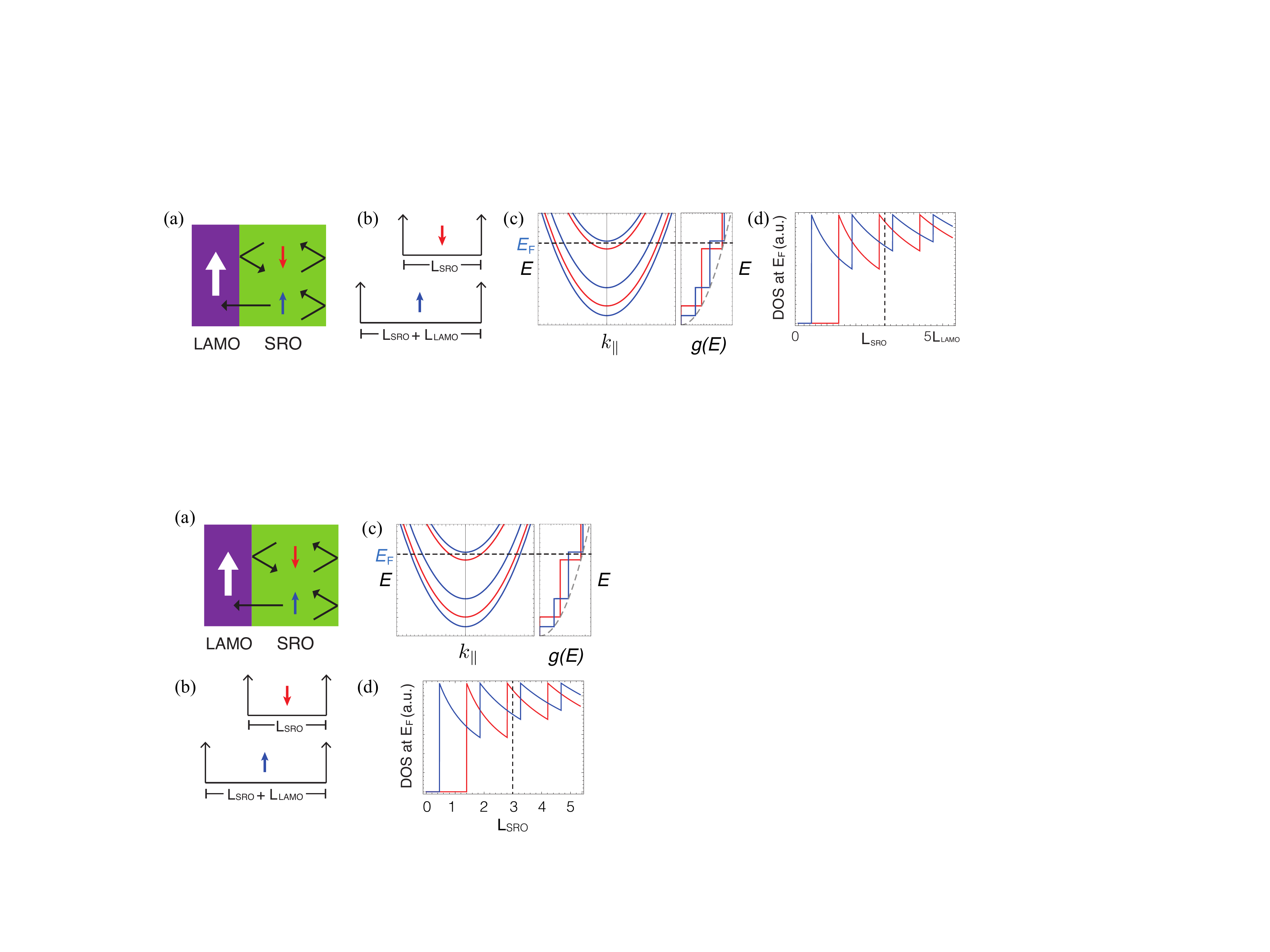}
\caption{Quantum well description of SRO-LAMO bilayers: (a) bilayer geometry, (b) Potentials seen by the two spin species; the up electrons see a wider quantum well, (c) The resulting band energies and DOS for the two spin species, and (d) DOS vs. width of SRO layer, keeping $L_{LAMO}$ fixed at unity. The DOS is shown for an arbitrarily chosen Fermi energy. }
\label{fig.QW_bilayer}
\end{figure}

Let us first consider SRO-LAMO multilayers as depicted in Fig.~\ref{fig.QW_multilayer}(a). Each SRO layer can now be thought of as a two-dimensional quantum well for down-electrons. 
 The boundary conditions are set by the adjacent LAMO layers -- up-electrons can diffuse freely into LAMO layers while down-electrons cannot. The down-spin electrons are localized within 2D quantum wells, while the up-electrons move freely within a 3D system. 
We find the density of states (DOS) of each spin species taking the electronic dispersion to be that of a free particle in a quantum well with the appropriate width. The up-electrons occupy states are described by a three-dimensional momentum quantum number. The down-electrons, however, have one momentum component ($k_\perp$) quantized due to the finite width of the 2D quantum wells. Their dispersion, as shown in Fig.~\ref{fig.QW_multilayer}(c), forms bands corresponding to each $k_\perp$ value. The DOS of both spin species is shown in Fig.~\ref{fig.QW_multilayer}(c) with the up-electrons always having higher DOS. As we are just above $T_c^{SRO}$, incipient ferromagnetic correlations arise inside each SRO layer. As SRO is a `minority-carrier' ferromagnet, the higher density of up-spin conduction electrons forces the magnetic moment of SRO to point down. This explains the antiferromagnetic coupling seen in LAMO-SRO multilayers.

The situation is more nuanced in LAMO-SRO bilayers, depicted in Fig.~\ref{fig.QW_bilayer}(a). In this case, both up and down electrons reside in 2D quantum wells; however, the widths of the two quantum wells are different as shown in Fig.~\ref{fig.QW_bilayer}(b). The resulting dispersion and DOS are shown in Fig.~\ref{fig.QW_bilayer}(c). Unlike the multilayer case, the nature of the dominant spin (that with higher DOS) depends on the precise value of the fermi energy.
If up-spin has higher DOS, we expect SRO to order with a moment pointing down and vice versa. This suggests a smoking gun signature of our conduction-electron driven mechanism. 
It is somewhat difficult to tune the fermi energy in experiments. However, it is relatively easy to make samples with different relative widths of the SRO and LAMO layers. Tuning the relative widths of the layers also changes the nature of the dominant spin as shown in Fig.~\ref{fig.QW_bilayer}(d). As a consequence, the ordered moment of SRO layers will switch direction. 
This suggests a smoking gun signature of our carrier-driven mechanism:
by varying the relative width of the SRO-LAMO layers, we can change the nature of the inter-layer coupling from antiferromagnetic to ferromagnetic. This is in sharp contrast to a superexchange picture in which the coupling only depends on the interface and is insensitive to the widths of the layers.

\paragraph{Discussion:}
We have proposed a new carrier-driven mechanism for inter-layer coupling in ferromagnetic heterostructures. 
We have illustrated this using SRO-LAMO bilayers and multilayer heterostructures as suitable test cases.
For the multilayer case, we have presented a theoretical microscopic model amenable to semi-classical simulations. The model qualitatively reproduces key experimental results without invoking interface-based mechanisms such as superexchange. This strongly supports our picture of carrier-driven coupling. 

We have presented a simple theoretical argument, modelling the LAMO and SRO layers as quantum-wells. Note that effects of quantum confinement have previously been reported in SRO slabs\cite{Chang2009}. Here, the density of states within the quantum wells shows spin dependence. Relying on the minority-carrier nature of SRO, we predict the direction of the ordered moment that will develop in SRO. Our results suggest that SRO-LAMO multilayers will always show antiferromagnetic coupling. However, in SRO-LAMO bilayers, we predict that the coupling can be tuned by varying the relative layer widths. This can be easily verified with current experimental setups. 

Our results show a deep connection between magnetic ordering and spin polarization of charge carriers. This suggests new avenues for manipulating spintronic devices. For instance, an ferromagnetic bilayer can be made antiferromagnetic by injection of a suitably polarized spin current. 
Alternatively, an antiferromagnetic bilayer can be thought of as a spin-polarized analogue of colossal magnetoresistance. At zero field, a large spin polarized current can flow. When a large enough magnetic field turns the coupling ferromagnetic, the spin polarization of currents is strongly diminished.
Conversely, heterostructure geometries can be tweaked to obtain suitably polarized charge carriers. They can be used as spin filters that can be tuned by an external magnetic field. 

\bibliographystyle{apsrev4-1}
\bibliography{LSMO_SRO,LSMO_SRO_SK}

\newpage
\appendix
\section{Supplementary Material: Semi-classical Monte Carlo simulations}
To solve the Hamiltonian given by Eq. \ref{eq.Hamiltonian} on this cluster, we treat the local moment as a classical spin. These spins provide a spin-dependent potential background for the conduction electrons. 
For a given local moment background, we diagonalize the single-electron problem and take states lying below the Fermi level to be filled. 
The Fermi level itself is chosen so as to give the correct number of electrons in the system. 
Our approach is a generalization of the classical Monte Carlo approach for spin systems: at each step, we propose a change in the local moment configuation. 
We rediagonalize the electron problem, find the Fermi level, and calculate the kinetic energy cost to the electrons. The change in total energy due to the proposed local moment 
update has two contributions: from the electron kinetic energy and from the Kondo coupling of conduction electrons to local moments. 
As in conventional classical Monte Carlo, we decide whether or not to accept this proposed change as follows: if 
the energy is lowered, we accept it, if energy increases, we accept the proposal with probability $\exp(-\Delta E/T)$.
Typically, for a given temperature, we make $\sim 10^3$ Monte Carlo updates until the system thermalizes. 
Subsequently, a similar number of Monte Carlo configurations are used for computing thermal and quantum averages.
This method has been extensively used in the study of models where some degrees of freedom can be approximated as classical variables\cite{Dagotto2001a,Nasu2014,Nasu2015}.

\end{document}